\documentclass[twocolumn,showpacs,preprintnumbers,prd,nofootinbib]{revtex4}
\usepackage{epsfig,amsmath,amssymb}

\pdfoutput=1

\newcommand{\beq}{\begin{equation}}
\newcommand{\eeq}{\end{equation}}
\newcommand{\bea}{\begin{eqnarray}}
\newcommand{\eea}{\end{eqnarray}}
\newcommand{\Fig}[1]{Fig.\,\ref{#1}}

\newcommand{\Eq}[1]{Eq.\,(\ref{#1})}
\newcommand{\eq}[1]{(\ref{#1})}
\newcommand{\Eqs}[1]{Eqs.\,(\ref{#1})}

\newcommand{\Tab}[1]{Tab.\,\ref{#1}}

\newcommand{\f}{\frac}

\newcommand{\sw}{s_{\scriptscriptstyle W}}
\newcommand{\cw}{c_{\scriptscriptstyle W}}
\newcommand{\sws}{s^2_{\scriptscriptstyle W}}
\newcommand{\cws}{c^2_{\scriptscriptstyle W}}

\newcommand{\MZ}{M_{\scriptscriptstyle Z}}

\newcommand{\MHpm}{M^\pm_{H}}

\newcommand{\mb}{m_b}
\newcommand{\mt}{m_t}

\newcommand{\mcha}{M^\pm_{{\tilde \chi_1}}}

\newcommand{\GF}{G_F}

\newcommand{\GeV}{{\rm GeV}}
\newcommand{\TeV}{{\rm TeV}}

\newcommand{\re}{{\rm Re}}

\def\unit{\leavevmode\hbox{\small1\kern-3.6pt\normalsize1}}
\newcommand{\eps}{\epsilon}

\newcommand{\sL}{{\scalebox{0.6}{$L$}}}

\newcommand{\Kpnn}{K^+ \to \pi^+ \nu \bar{\nu}}
\newcommand{\KLnn}{K_L \to \pi^0 \nu \bar{\nu}}

\newcommand{\KLmm}{K_L \to \mu^+ \mu^-}

\newcommand{\BXsga}{\bar{B} \to X_s \gamma}

\newcommand{\BXsll}{\bar{B} \to X_s l^+ l^-}

\newcommand{\BXdnn}{\bar{B} \to X_d \nu \bar{\nu}}
\newcommand{\BXsnn}{\bar{B} \to X_s \nu \bar{\nu}}

\newcommand{\Bdmm}{B_d \to \mu^+ \mu^-}
\newcommand{\Bsmm}{B_s \to \mu^+ \mu^-}

\newcommand{\BRKp}{{\cal B} (\Kpnn)}
\newcommand{\BRKL}{{\cal B} (\KLnn)}
\newcommand{\BRKm}{{\cal B} (\KLmm)_{\rm SD}}
\newcommand{\BRXd}{{\cal B} (\BXdnn)}
\newcommand{\BRXs}{{\cal B} (\BXsnn)}
\newcommand{\BRBd}{{\cal B} (\Bdmm)}
\newcommand{\BRBs}{{\cal B} (\Bsmm)}
\newcommand{\BRga}{{\cal B} (\BXsga)}
\newcommand{\BRll}{{\cal B} (\BXsll)}

\newcommand{\btosgamma}{b \to s \gamma}

\newcommand{\Ztobb}{Z \to b \bar{b}}

\newcommand{\Ztodjdi}{Z \to d^j \bar{d}^i}
\newcommand{\Zbb}{Z b_\sL \bar{b}_\sL}
\newcommand{\Zdjdi}{Z d^j_\sL \bar{d^i_\sL}}

\newcommand{\C}{C}

\newcommand{\Rb}{R_b^0}
\newcommand{\Ab}{{\cal A}_b}
\newcommand{\AFB}{A_{\rm FB}^{0, b}}

\newcommand{\CP}{C\hspace{-0.25mm}P}

\newcommand{\mysigma}{\hspace{0.4mm} \sigma}

\newcommand{\etal}{{\it et al}.}

\begin{document}

\allowdisplaybreaks

\preprint{ZU-TH 10/07} 

\title{ \boldmath \hspace{5mm}
  \mbox{\scalebox{0.75}{\includegraphics{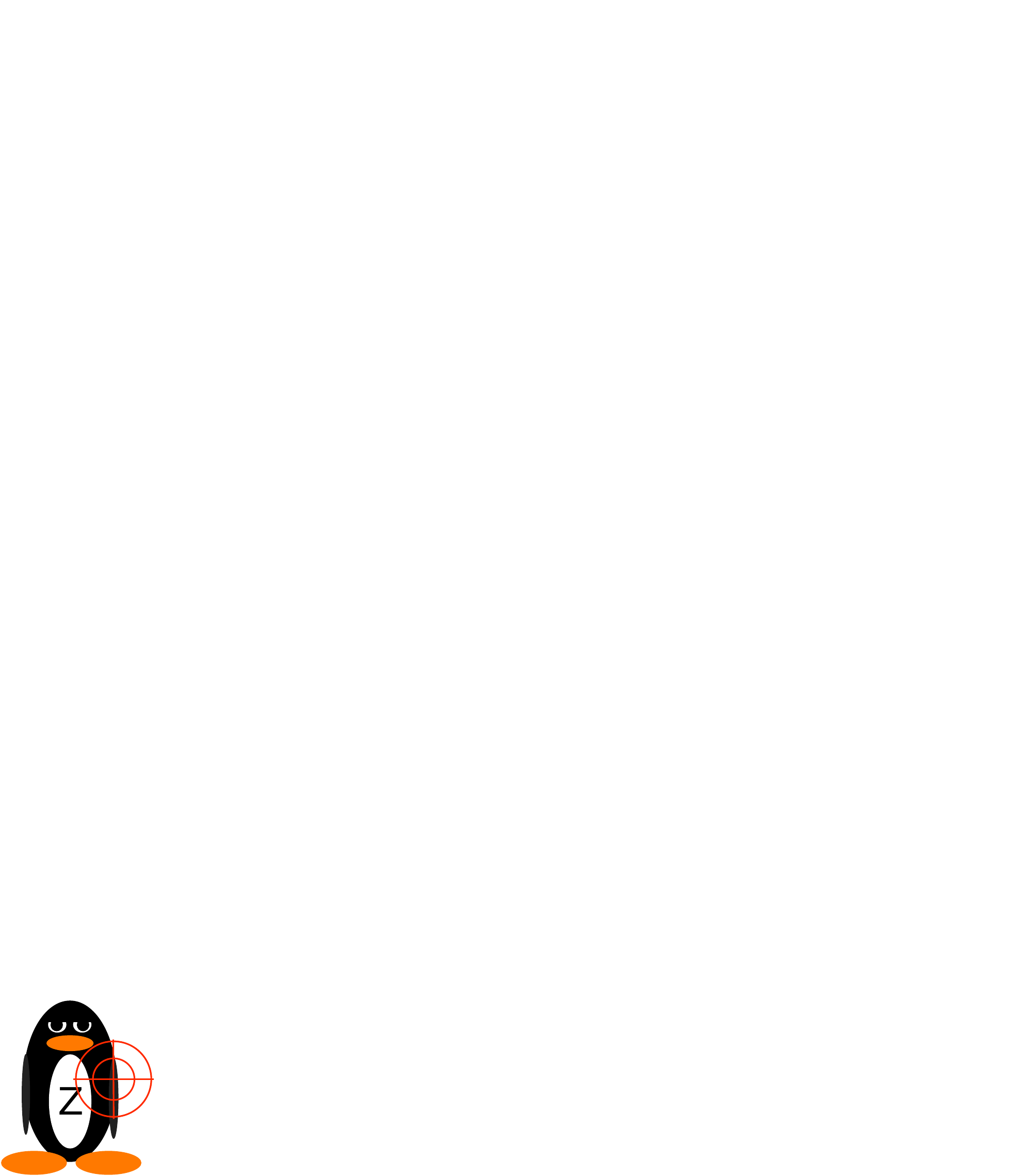}} \hspace{-3.15cm}
    How To Kill a Penguin\footnote[2]{Talk given at DIS 2007,
      M\"unchen, Germany, April 15--20, 2007.}}  \unboldmath }

\author{Ulrich~Haisch} 

\affiliation{
Institut f\"ur Theoretische Physik, Universit\"at Z\"urich,
CH-8057 Z\"urich, Switzerland 
}

\date{\today}

\begin{abstract}
\noindent
Within constrained minimal-flavor-violation the large destructive
flavor-changing $Z$-penguin managed to survive eradication so far. We
give a incisive description of how to kill it using the precision
measurements of the $\Ztobb$ pseudo observables. The derived stringent
range for the non-standard contribution to the universal Inami-Lim
function $\C$ leads to tight two-sided limits for the branching ratios
of all $Z$-penguin dominated flavor-changing $K$- and $B$-decays.
\end{abstract}


\maketitle

\section{Introduction}
\label{sec:introduction}

The effects of new heavy particles appearing in extensions of the
standard model (SM) can be accounted for at low energies in terms of
effective operators. The unprecedented accuracy reached by the
electroweak (EW) precision measurements performed at the high-energy
colliders LEP and SLC impose stringent constraints on the coefficients
of the operators entering the EW sector. Other severe constraints came
in recent years from the BaBar, Belle, CDF, and D\O\ experiments and
concern extra sources of flavor and $C\hspace{-0.5mm}P$ violation that
represent a generic problem in many scenarios of new physics (NP). The
most pessimistic but experimentally well supported solution to the
flavor puzzle is to assume that all flavor and $C\hspace{-0.5mm}P$
violation is governed by the known structure of the SM Yukawa
interactions. In these minimal-flavor-violating (MFV)
\cite{Chivukula:1987py, MFV, D'Ambrosio:2002ex} models correlations
between certain flavor diagonal high-energy and flavor off-diagonal
low-energy observables exist since, by construction, NP couples
dominantly to the third generation. In order to simplify matters, we
restrict ourselves in the following to the class of constrained MFV
(CMFV) \cite{Blanke:2006ig} models, i.e., scenarios that involve only SM
operators, and thus consider just left-handed currents.

\section{General considerations}
\label{sec:general}

That new interactions unique to the third generation can lead to an
intimate relation between the non-universal $\Zbb$ and the flavor
non-diagonal $\Zdjdi$ vertices has been shown recently in
\cite{Haisch:2007ia}. Whereas the former structure is probed by the
ratio of the $Z$-boson decay width into bottom quarks and the total
hadronic width, $\Rb$, the bottom quark asymmetry parameter, $\Ab$,
and the forward-backward asymmetry for bottom quarks, $\AFB$, the
latter ones appear in many $K$- and $B$-decays.

In the effective field theory framework of MFV
\cite{D'Ambrosio:2002ex}, one can easily see how the $\Zbb$ and
$\Zdjdi$ operators are linked together. The only relevant
dimension-six contributions compatible with the flavor group of MFV
stem from the $SU(2) \times U(1)$ invariant operators
\beq \label{eq:zoperators}
\begin{aligned}
  {\cal O}_1 & = i \left( {\bar Q}_L Y_U Y_U^\dagger \gamma_\mu Q_L
  \right) \phi^\dagger D^\mu \phi \, , \\
  {\cal O}_2 & = i \left( {\bar Q}_L Y_U Y_U^\dagger \tau^a \gamma_\mu
    Q_L\right) \phi^\dagger \tau^a D^\mu \phi \, ,
\end{aligned}
\eeq 
that are built out of the quark doublets $Q_L$, the Higgs field
$\phi$, the up-type Yukawa matrices $Y_U$, and the $SU(2)$ generators
$\tau^a$. After EW symmetry breaking, ${\cal O}_{1, 2}$ are
responsible for both the effective $\Zbb$ and $\Zdjdi$ vertex. Since
all up-type quark Yukawa couplings except the one of the top, $y_t$,
are small, one has $(Y_U Y_U^\dagger)_{ji} \sim y_t^2 V_{tj}^\ast
V_{ti}$ and only this contribution matters in \Eq{eq:zoperators}.

Within the SM the Feynman diagrams responsible for the enhanced top
correction to the $\Zbb$ coupling also generate the $\Zdjdi$
operators. In fact, in the limit of infinite top quark mass the
corresponding amplitudes are up to Cabibbo-Kobayashi-Maskawa (CKM)
factors identical. Yet there is a important difference between
them. While for the physical $\Ztobb$ decay the diagrams are evaluated
on-shell, in the case of the low-energy $\Ztodjdi$ transitions the
amplitudes are Taylor-expanded up to zeroth order in the external
momenta. As far as the momentum of the $Z$-boson is concerned the two
cases correspond to $q^2 = \MZ^2$ and $q^2 = 0$.

\begin{figure}[t!]
\begin{center}
\begin{picture}(245,145)(0,0)
\put(0,0){\makebox{\hspace{+0mm} \scalebox{0.775}{\hspace{2mm} \includegraphics{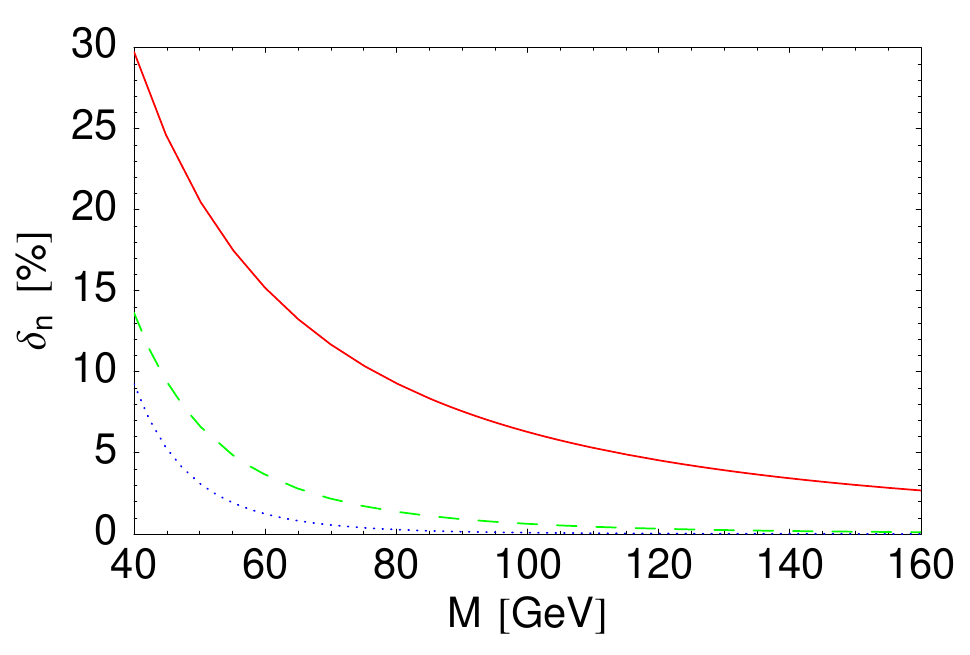}}}}
\put(95,60){\makebox{\hspace{+0mm} \scalebox{0.375}{\includegraphics{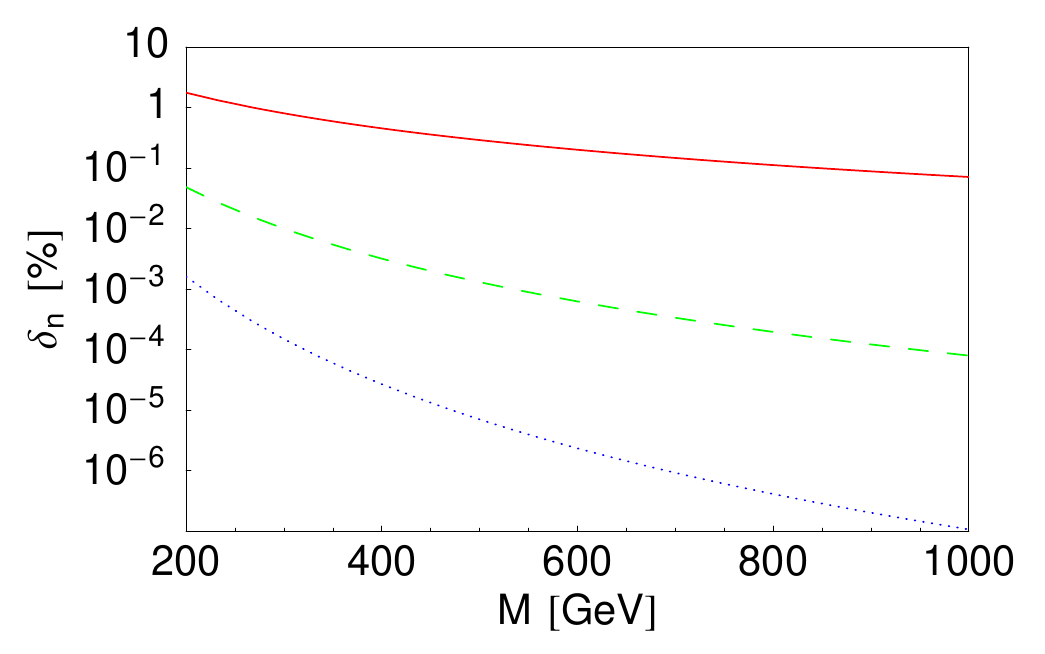}}}}
\end{picture}
\end{center}
\vspace{-6mm}
\caption{\sf Relative deviations $\delta_n$ as a function of $M$. The
  solid, dashed, and dotted curve correspond to $n = 1,2,$ and $3$,
  respectively. See text for details.}
\label{fig:sme}
\end{figure}

The general features of the small momentum expansion of the one-loop
$\Ztobb$ vertex can be nicely illustrated with the following simple
but educated example. Consider the scalar integral
\beq \label{eq:c0}
C_0 = \f{m_3^2}{i \pi^2} \int \! \f{d^4 l}{D_1 D_2 D_3} \, ,
\hspace{5mm} D_i = (l + p_i)^2 - m_i^2 \, ,
\eeq
with $p_3 = 0$. Note that we have set the space-time dimension to four
since the integral is finite and assumed without loss of generality
$m_3 \neq 0$.

In the limit of vanishing bottom quark mass one has for the
corresponding momenta $p^2_1 = p^2_2 = 0$. The small momentum expansion
of the scalar integral $C_0$ then takes the form  
\beq \label{eq:sme}
C_0 = \sum_{n = 0}^\infty a_n \left ( \f{q^2}{m_3^2} \right )^n \, ,
\eeq
with $q^2 = (p_1 - p_2)^2 = -2 \hspace{0.2mm}  p_1 \! \cdot \! p_2$. The
expansion coefficients $a_n$ are given by \cite{Fleischer:1994ef}
\beq \label{eq:an}
a_n = \f{(-1)^n}{(n + 1)!} \sum_{l = 0}^n \begin{pmatrix} n \\
  l \end{pmatrix} \f{x_1^l}{l!}  \f{\partial^l}{\partial x_1^l}
\f{\partial^n}{\partial x_2^n} g(x_1, x_2) \, ,
\eeq
where
\beq \label{eq:gx1x2}
g(x_1, x_2) = \f{1}{x_1 - x_2} \left ( \f{x_1 \ln x_1}{1 - x_1} -
  \f{x_2 \ln x_2}{1 - x_2} \right ) \, ,
\eeq
and $x_i = m_i^2/m_3^2$. Notice that in order to properly generate the
expansion coefficients $a_n$ one has to keep $x_1$ and $x_2$ different
even in the zero or equal mass case. The corresponding limits can only
be taken at the end.

To illustrate the convergence behavior of the small momentum expansion
of the scalar integral in \Eq{eq:sme} for on-shell kinematics, we
confine ourselves to the simplified case $m_1 = m_2 = M$ and $m_3 =
\mt$. We define
\beq \label{eq:deltan}
\delta_n = a_n \left ( \f{\MZ^2}{\mt^2} \right )^n \left ( \sum_{l =
    0}^{n - 1} a_l \left ( \f{\MZ^2}{\mt^2} \right )^l \right )^{-1}
\, ,
\eeq
for $n = 1, 2, \ldots \,$. The $M$-dependence of the relative
deviations $\delta_n$ is displayed in \Fig{fig:sme}. We see that while
for $M \lesssim 50 \, \GeV$ higher order terms in the small momentum
expansion have to be included in order to approximate the exact
on-shell result accurately, in the case of $M \gtrsim 150 \, \GeV$ the
first correction is small and higher order terms are negligible. For
the two reference scales $M = \{ 80, 250 \} \, \GeV$ one finds for the
first three relative deviations $\delta_n$ numerically $+9.3 \%$,
$+1.4 \%$, and $+0.3 \%$, and $+1.1 \%$, $+0.02 \%$, $+0.00004 \%$,
respectively.

Of course the two reference points $M = \{ 80, 250 \} \, \GeV$ have
been picked for a reason. While the former describes the situation in
the SM, i.e., the exchange of two pseudo Goldstone bosons and a top
quark, the latter presents a possible NP contribution involving
besides the top, two heavy scalars. The above example indicates that
the differences between the $\Zbb$ form factor evaluated on-shell and
at zero external momenta are in general much less pronounced in models
with new heavy degrees of freedom than in the SM. Given that this
difference amounts to around $-30 \%$ in the SM \cite{zbb}, it is
suggestive to assume that the scaling of NP contributions to the
non-universal $\Zbb$ vertex is in general under $\pm 10 \%$. This
model-independent conclusion is well supported by the results of the
calculations of the one-loop $\Zbb$ vertices in popular CMFV models
presented in \cite{Haisch:2007ia}.

\section{Model calculations}
\label{sec:calculations}

\begin{figure}[!t]
\begin{center}
\hspace{-10mm}
\scalebox{0.75}{\hspace{10mm} \includegraphics{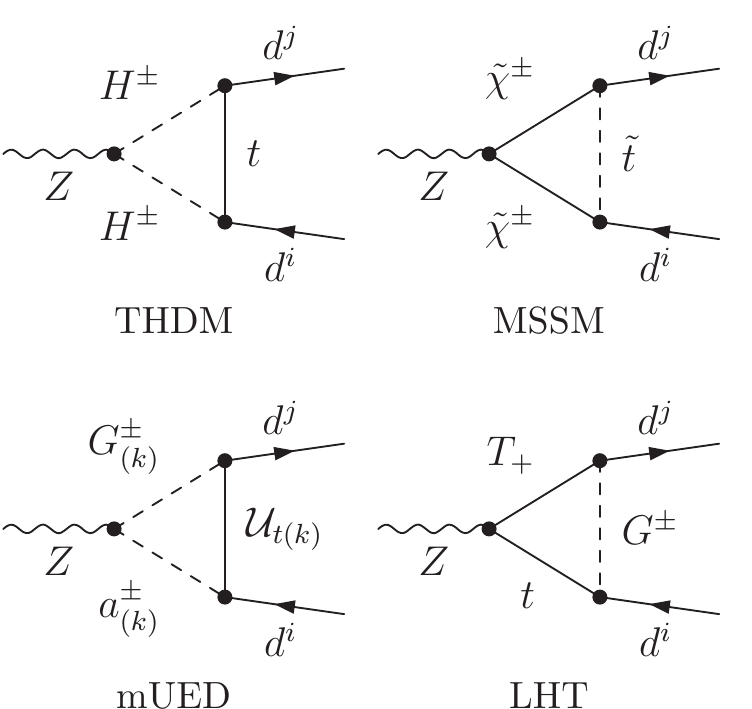}}
\end{center}
\vspace{-4mm}
\caption{\sf Examples of one-loop vertex diagrams that result in a  
  non-universal correction to the $\Ztodjdi$ transition in assorted 
  NP scenarios with CMFV. See text for details. }  
\label{fig:cmfv}
\end{figure}

The above considerations can be corroborated in another, yet
model-dependent way by calculating explicitly the difference between
the value of the $\Zdjdi$ vertex form factor evaluated on-shell and at
zero external momenta. In \cite{Haisch:2007ia} this has been done in
four of the most popular, consistent, and phenomenologically viable
scenarios of CMFV, i.e., the two-Higgs-doublet model (THDM) type I and
II, the minimal-supersymmetric SM (MSSM) with MFV \cite{MFV}, all for
small $\tan \beta$, the minimal universal extra dimension (mUED) model
\cite{Appelquist:2000nn}, and the littlest Higgs model
\cite{Arkani-Hamed:2002qy} with $T$-parity (LHT) \cite{tparity} and
degenerate mirror fermions \cite{Low:2004xc}. Examples of diagrams
that contribute to the $\Ztodjdi$ transition in these models can be
seen in \Fig{fig:cmfv}. In the following we will briefly summarize the
most important findings of \cite{Haisch:2007ia}.

\begin{figure}[!t]
\begin{center}
\makebox{\hspace{+0mm} \scalebox{0.7}{\includegraphics{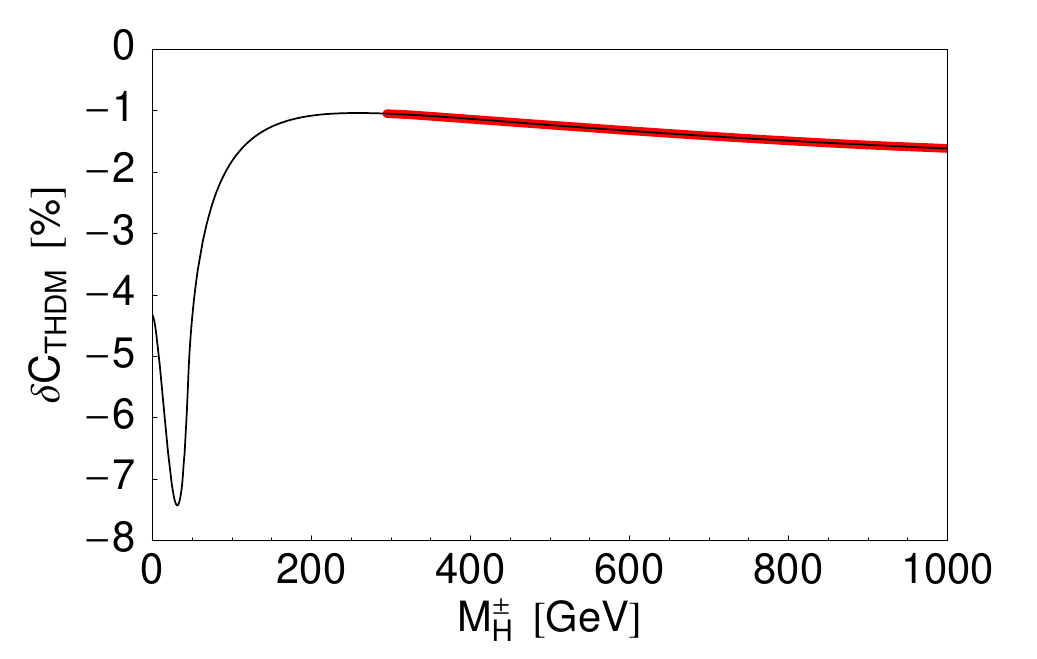}}}

\vspace{-2mm}

\makebox{\hspace{-3mm} \scalebox{0.714}{\includegraphics{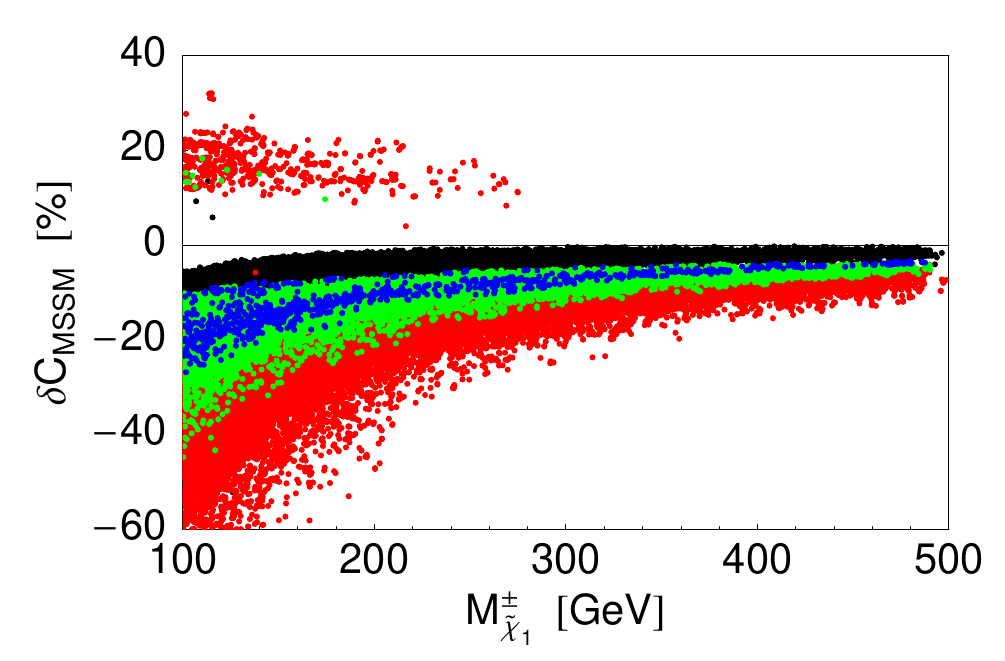}}}

\vspace{-4mm}

\makebox{\hspace{-2mm} \scalebox{0.7}{\includegraphics{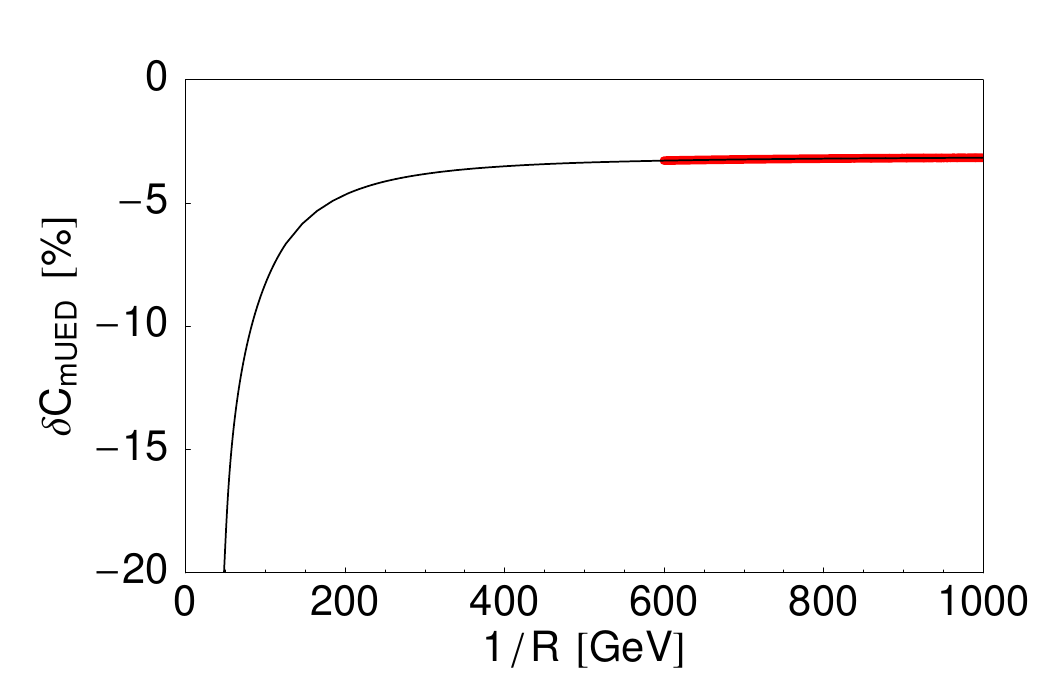}}}

\vspace{-2mm}

\makebox{\hspace{-6mm} \scalebox{0.7}{\includegraphics{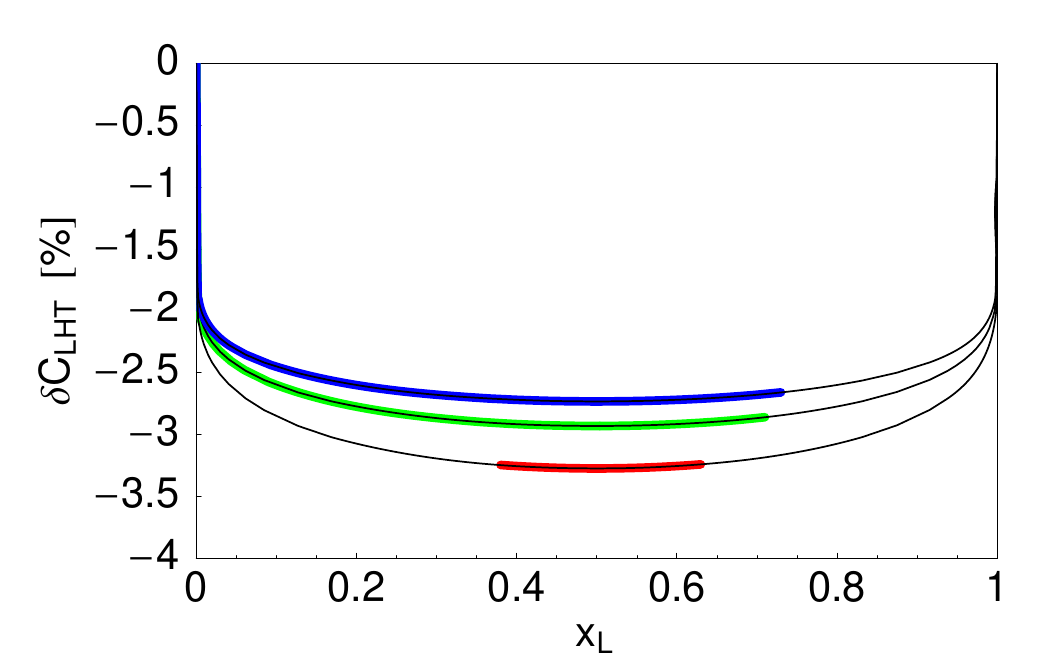}}}
\end{center}
\vspace{-6mm}
\caption{\sf Relative difference $\delta C_{\rm NP}$ in the THDMs, the
  MSSM, the mUED, and the LHT model as a function of $\MHpm$, $\mcha$,
  $1/R$, and $x_L$. Regions in the $\mcha$--$\, \delta C_{\rm MSSM}$
  plane where $| \re \, C_{\rm MSSM} (q^2 = 0)|$ amounts to at least
  $2 \%$, $4 \%$, and $6 \%$ of $| \re \, C_{\rm SM} (q^2 = 0) |$ are
  indicated by the red (gray), green (light gray), and blue (dark
  gray) points, respectively. In the case of the LHT model the shown
  curves correspond, from bottom to top, to the values $f = 1, 1.5$,
  and $2 \, \TeV$ of the symmetry breaking scale. See text for
  details.}
\label{fig:scalings}
\end{figure}

In the limit of vanishing bottom quark mass, possible non-universal NP
contributions to the renormalized off-shell $\Zdjdi$ vertex can be
written as
\beq \label{eq:zdjdi}
\Gamma_{ji}^{\rm NP} = \f{\GF}{\sqrt{2}} \f{e}{\pi^2} \MZ^2
\f{\cw}{\sw} V_{tj}^\ast V_{ti} C_{\rm NP} (q^2) \bar{d^j}_\sL
\gamma_\mu {d^i}_\sL Z^\mu \, ,   
\eeq
where $i = j = b$ and $i \neq j$ in the flavor diagonal and
off-diagonal case. $\GF$, $e$, $\sw$, and $\cw$ denote the Fermi
constant, the electromagnetic coupling constant, the sine and cosine
of the weak mixing angle, respectively, while $V_{ij}$ are the
corresponding CKM matrix elements.

As a measure of the relative difference between the complex valued
form factor $C_{\rm NP} (q^2)$ evaluated on-shell and at zero momentum 
we introduce  
\beq \label{eq:dcnp}
\delta C_{\rm NP} = 1 - \f{\re \, C_{\rm NP} (q^2 = 0)}{\re \, C_{\rm
    NP} (q^2 = \MZ^2)} \, .
\eeq  

\begin{figure}[!t]
\vspace{-5mm}
\scalebox{0.4}{\includegraphics{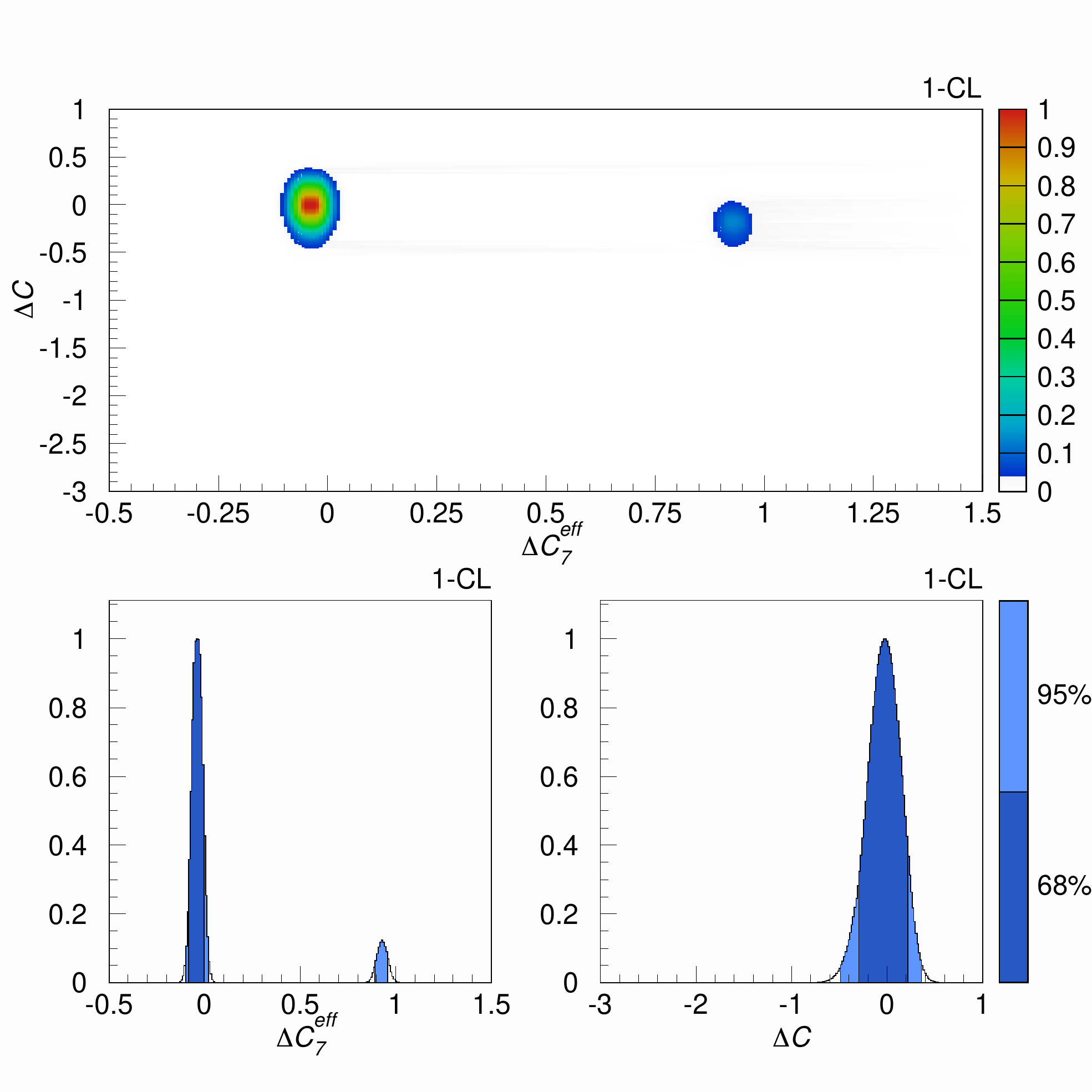}}
\vspace{-4mm}
\caption{\sf Constraints on $\Delta C_7^{\rm eff}$ and $\Delta C$
  within CMFV that follow from a combination of the $\Ztobb$ POs with
  the measurements of $\BXsga$ and $\BXsll$. The colors encode the
  frequentist $1 - {\rm CL}$ level and the corresponding $68 \%$ and
  $95 \%$ probability regions as indicated by the bars on the right
  side of the panels. See text for details.}
\label{fig:dcdc7effnow}
\end{figure}

The dependence of $\delta C_{\rm NP}$ on the charged Higgs mass
$\MHpm$, the lighter chargino mass $\mcha$, the compactification scale
$1/R$, and $x_L$ which parameterizes the mass of the heavy top $T_+$ is
illustrated in \Fig{fig:scalings}. The allowed parameter regions after
applying experimental and theoretical constraints are indicated by the
colored (grayish) bands and points.

In the THDMs, the mUED, and the CMFV version of the LHT model the
maximal allowed suppressions of $\re \, C_{\rm NP} (q^2 = \MZ^2)$ with
respect to $\re \, C_{\rm NP} (q^2 = 0)$ amounts to less than $2 \%$,
$5 \%$, and $4 \%$, respectively. This feature confirms the general
argument presented in the last section. The situation is less
favorable in the case of the CMFV MSSM, since $\delta C_{\rm MSSM}$
frequently turns out to be larger than one would expected on the basis
of the model-independent considerations if the masses of the lighter
chargino and stop both lie in the hundred $\GeV$ range. However, the
large deviation $\delta C_{\rm MSSM}$ are ultimately no cause of
concern, because $|\re \, C_{\rm MSSM} (q^2 = 0)/\re \, C_{\rm SM}
(q^2 = 0)|$ itself is always below $10 \%$. In consequence, the
model-independent bounds on the NP contribution to the universal
$Z$-penguin function that will be derived in the next section do hold
in the case of the CMFV MSSM. More details on the phenomenological
analysis of $\delta C_{\rm NP}$ in the THDMs, the CMFV MSSM, the mUED,
and the LHT model including the analytic expressions for the form
factors $C_{\rm NP} (q^2)$ can be found in the recent article
\cite{Haisch:2007ia}.

\begin{widetext}
\begin{center}
\begin{table}[!t]
\caption{\sf Bounds for various rare decays in CMFV models at $95 \%$
  probability, the corresponding values in the SM at $68 \%$ and $95
  \% \, {\rm CL}$, and the available experimental information. See
  text for details.}    
\vspace{1mm} 
\begin{center}
\begin{tabular}{c@{\hspace{2.5mm}}cccc}
\hline \hline \\[-4.5mm]
Observable & CMFV ($95 \% \, {\rm CL}$) & \hspace{0mm} SM ($68 \% \,
{\rm CL}$) & \hspace{0mm} SM ($95 \% \, {\rm CL}$) & \hspace{0mm}
Experiment \\[0.5mm] 
\hline 
$\BRKp \times 10^{11}$ & $[4.29, 10.72]$ & \hspace{0mm} $7.15 \pm 1.28$ &  
\hspace{0mm} $[5.40, 9.11]$ & \hspace{0mm} $\left ( 14.7^{+13.0}_{-8.9}
\right )$ \cite{kp} \\
$\BRKL \times 10^{11}$ & $[1.55, 4.38]$ & \hspace{0mm} $2.79 \pm 0.31$
& \hspace{0mm} $[2.21, 3.45]$ & \hspace{0mm} $< 2.1 \times 10^4 \;\;
(90 \% \, \text{CL})$ \cite{Ahn:2006uf} \\
$\BRKm \times 10^9$ & $[0.30, 1.22]$ & \hspace{0mm} $0.70 \pm 0.11$ &
\hspace{0mm} $[0.54, 0.88]$ & \hspace{0mm} -- \\
$\BRXd \times 10^6$ & $[0.77, 2.00]$ & \hspace{0mm} $1.34 \pm 0.05$ &
\hspace{0mm} $[1.24, 1.45]$ & \hspace{0mm} -- \\
$\BRXs \times 10^5$ & $[1.88, 4.86]$ & \hspace{0mm} $3.27 \pm 0.11$ &
\hspace{0mm} $[3.06, 3.48]$ & \hspace{0mm} $< 64 \;\; (90 \% \,
\text{CL})$ \cite{Barate:2000rc} \\ 
$\BRBd \times 10^{10}$ & $[0.36, 2.03]$ & \hspace{0mm} $1.06 \pm 0.16$ &
\hspace{0mm} $[0.87, 1.27]$ & \hspace{0mm} $< 3.0 \times 10^2 \;\; (95
\% \, \text{CL})$ \cite{Bernhard:2006fa} \\  
$\BRBs \times 10^9$ & $[1.17, 6.67]$ & \hspace{0mm} $3.51 \pm 0.50$ &
\hspace{0mm} $[2.92, 4.13]$ & \hspace{0mm} $< 9.3 \times 10^1
\;\; (95 \% \, \text{CL})$ \cite{Sanchez:2007ew} \\[1mm] 
\hline \hline
\end{tabular}
\end{center}
\label{tab:brs}
\end{table}
\end{center}
\end{widetext}

\section{Numerical analysis}
\label{sec:numerics}

Using the technique of epsilon parameters a model-independent
numerical analysis of $\Delta C = \re \, C (q^2 = 0) - \re \, C_{\rm
  SM} (q^2 = 0)$ is a back-on-the-envelope calculation. The variation
$\eps_b^{\rm NP} = \eps_b - \eps_b^{\rm SM}$ arising from NP
contributions to $\Zbb$ can be defined through the inclusive partial
width of $\Ztobb$ as follows \cite{epsilonb}
\beq \label{eq:Gammabb}
\Gamma^{\rm NP}_{b b} = (\sqrt{2} \GF \MZ^2)^{\f{1}{2}} \, \Big (
g_V^b (\bar b \gamma_\mu b) - g_A^b (\bar b \gamma_\mu \gamma_5 b)
\Big ) Z^\mu \, ,
\eeq
where
\beq \label{eq:gVgA}
\f{g_V^b}{g_A^b} = \left ( 1 + \f{4 \sws}{3} + \eps_b^{\rm NP} \right
) \f{g_A^d}{g_A^b} \, , \hspace{2.5mm} g_A^b = (1 + \eps_b^{\rm NP}) \,
g_A^d \, . ~
\eeq
From \Eqs{eq:zdjdi}, \eq{eq:dcnp}, and \eq{eq:Gammabb} one obtains 
\beq \label{eq:DeltaCformulas}
\Delta C = -\f{\pi^2}{\sqrt{2} \GF \MZ^2 \cws} (1 + \delta C_{\rm NP}) 
\, \eps_b^{\rm NP} \, . 
\eeq
By combining experimental \cite{ewpm} and theoretical uncertainties
\cite{zfitter} in $\eps_b$ and $\eps_b^{\rm SM}$ linearly one finds
\beq \label{eq:epsbNP}
\eps_b^{\rm NP} = (0.4 \pm 2.5) \times 10^{-3} \, . 
\eeq
Assuming $\delta C_{\rm NP} = \pm 0.1$ one then arrives at
\beq \label{eq:DeltaC}
\Delta C = -0.04 \pm 0.26 \, , 
\eeq
which implies that large negative contributions that would reverse the
sign of the SM $Z$-penguin amplitude are highly disfavored in CMFV
scenarios due to the strong constraint from $\Rb$
\cite{Haisch:2007ia}. Interestingly, such a conclusion cannot be drawn
by considering only flavor constraints \cite{Bobeth:2005ck}, since a
combination of $\BRga$, $\BRll$, and $\BRKp$ does not allow to
distinguish the SM solution $\Delta C = 0$ from the wrong-sign case
$\Delta C \approx -2$ at present. 

The result in \Eq{eq:DeltaC} agrees amazingly well with the numbers of
a thorough global fit to the POs $\Rb$, $\Ab$, and $\AFB$ \cite{ewpm}
and the measured $\BXsga$ \cite{bsgamma} and $\BXsll$ \cite{bxsll} BRs
obtained by employing customized versions of the {\tt ZFITTER}
\cite{zfitter} and the CKMfitter package
\cite{Charles:2004jd}. Neglecting contributions from EW boxes these
bounds read \cite{Haisch:2007ia}
\beq \label{eq:dcsb0}
\begin{aligned}
\Delta C & = -0.026 \pm 0.264 & \!\! (68 \% \, {\rm CL}) \, , \\
\Delta C & = [-0.483, 0.368] & \!\! (95 \% \, {\rm CL}) \, . 
\end{aligned}
\eeq 
The constraint on $\Delta C$ within CMFV following from the
simultaneous use of $\Rb$, $\Ab$, $\AFB$, $\BRga$, and $\BRll$ can be
seen in \Fig{fig:dcdc7effnow}. 

One can also infer from this figure that two regions, resembling the
two possible signs of the amplitude ${\cal A} (\btosgamma) \propto
C_7^{\rm eff} (\mb)$, satisfy all existing experimental bounds. The
best fit value for $\Delta C_7^{\rm eff} = C_7^{\rm eff} (\mb) - C_{7
  \, {\rm SM}}^{\rm eff} (\mb)$ is very close to the SM point residing
in the origin, while the wrong-sign solution located on the right is
highly disfavored, as it corresponds to a $\BRll$ value considerably
higher than the measurements \cite{Gambino:2004mv}. The corresponding
limits are \cite{Haisch:2007ia}
\bea \label{eq:dc7effsb0}
\begin{gathered}
\Delta C_7^{\rm eff} = -0.039 \pm 0.043 \;\; (68 \% \, {\rm CL})
\, , \hspace{2mm} \\ 
\Delta C_7^{\rm eff} = [-0.104, 0.026] \, \cup \, [0.890, 0.968] 
\;\;  (95 \% \, {\rm CL}) \, . \hspace{2mm}  
\end{gathered} 
\eea 
Similar bounds have been presented previously in
\cite{Bobeth:2005ck}. Notice that since the SM prediction of $\BRga$
\cite{bsg} is now lower than the experimental world average by $1.2
\mysigma$, extensions of the SM that predict a suppression of the
$\btosgamma$ amplitude are strongly constrained. In particular, even
the SM point $\Delta C_7^{\rm eff} = 0$ is almost disfavored at $68 \%
\, {\rm CL}$ by the global fit.

The stringent bound on the NP contribution $\Delta C$ given in
\Eq{eq:dcsb0} translates into tight two-sided limits for the BRs of
all $Z$-penguin dominated flavor-changing $K$- and $B$-decays as shown
in \Tab{tab:brs}. A strong violation of any of the bounds by future
measurements will imply a failure of the CMFV assumption, signaling
either the presence of new effective operators and/or new flavor and
$\CP$ violation. A way to evade the given limits is the presence of
sizable corrections $\delta C_{\rm NP}$ and/or box
contributions. While these possibilities cannot be fully excluded,
general arguments and explicit calculations indicate that they are
both difficult to realize in the CMFV framework.

\section{Conclusions}
\label{sec:conclusions}

\begin{center}
  \scalebox{0.75}{\includegraphics{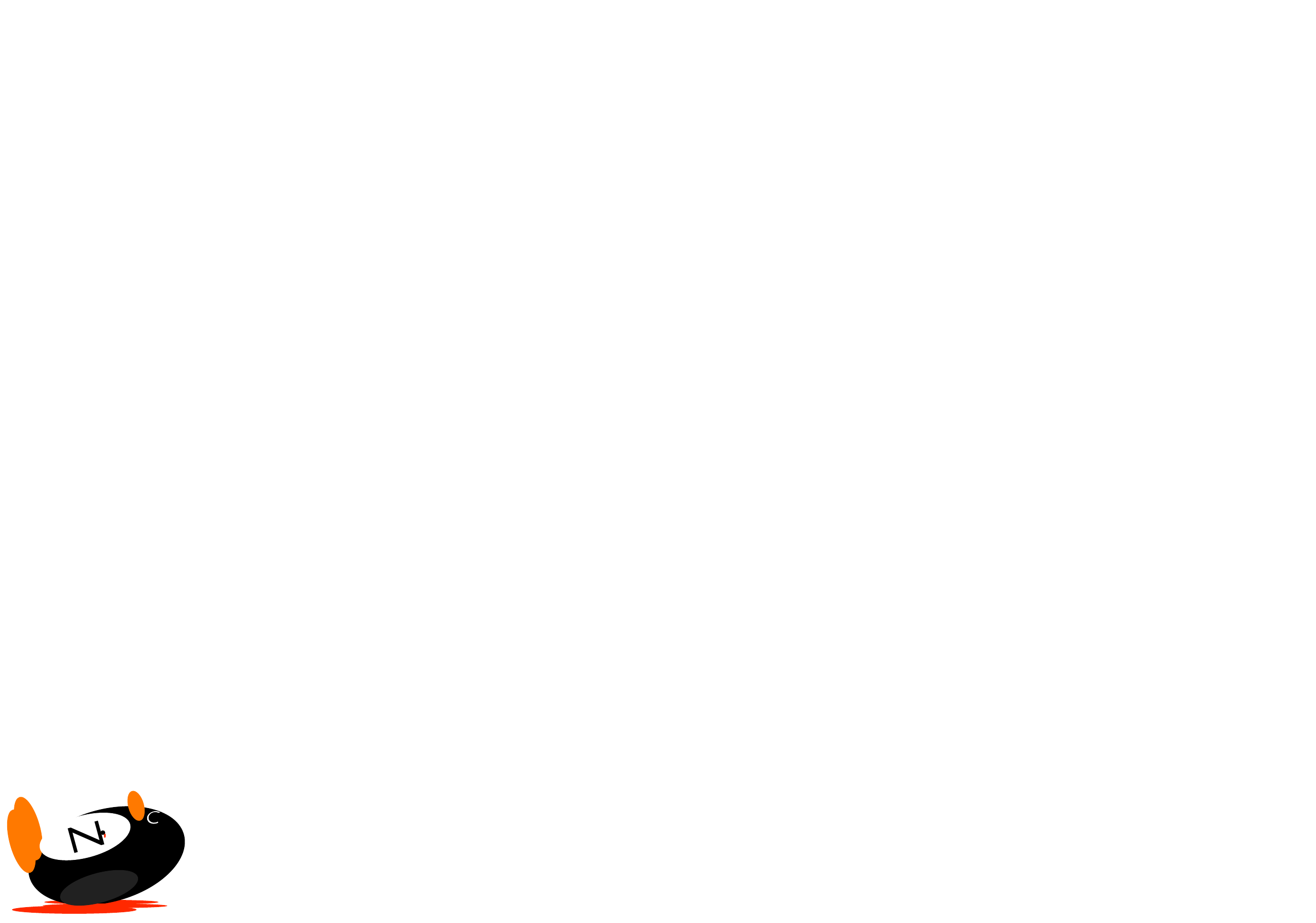}} \newline R.I.P. large
  destructive CMFV $Z$-penguin!
\end{center}

\acknowledgments{I am grateful to A.~Weiler for fruitful
  collaboration, valuable comments on the manuscript, and technical
  support. This work has been supported by the Schweizer Nationalfonds.}

\end{document}